\documentclass[twocolumn,english,aps,prb,floats,10pt]{revtex4-2}

\usepackage[T1]{fontenc}
\usepackage[utf8]{inputenc}
\usepackage{amsmath,amssymb,graphicx,siunitx}
\usepackage[unicode=true,bookmarks=false,breaklinks=false,pdfborder={0 0 1},backref=section,colorlinks=false]
{hyperref}

\DeclareSIUnit\angstrom{\text{\AA}}
\newcommand{\s}[1]{{\boldsymbol S}_{#1}}
\newcommand{\st}[2]{\tilde{\boldsymbol S}^{(#2)}_{#1}}
\newcommand{\heff}[1]{{\boldsymbol H}_{\mathrm{eff},#1}}
\newcommand{\hefft}[2]{\tilde{\boldsymbol H}^{(#2)}_{\mathrm{eff},#1}}
\newcommand{\nt}[2]{\tilde{\mathbf n}^{(#2)}_{#1}}
\newcommand{\om}{\boldsymbol\Omega}
\newcommand{\omt}[2]{\tilde{\om}^{(#2)}_{#1}}
\newcommand{\bv}[1]{\mathbf{#1}}
\newcommand{\bvt}[2]{\tilde{\mathbf{#1}}^{(#2)}}

\begin{document}

\begin{abstract}
We demonstrate that moving edge dislocations can induce the reversal of magnetization in a ferromagnetic film due to the Barnett effect. The dynamics of magnetization is studied numerically within a discretized Landau-Lifshitz equation on a hexagonal lattice containing over $10^5$ sites. Local coordinate frames coupled to the crystallographic axes for each spin are used together with the laboratory coordinate frame. The parameters of a hexagonal close-packed cobalt lattice have been chosen for illustration. The magnetization reversal from a metastable initial state created by the external magnetic field occurs on a time scale of a few picoseconds. Our results imply that fast local elastic twists generated by moving dislocations serve as an important mechanism of magnetization dynamics in solids subjected to a mechanical stress.
\end{abstract}

\title{Dislocation-induced magnetization reversal in a ferromagnetic film}
\author{ Jorge F. Soriano and Eugene M. Chudnovsky}
\affiliation{Physics Department, Herbert H. Lehman College and Graduate School, The City University of New York, 250 Bedford Park Boulevard West, Bronx, New York 10468-1589, USA }
\date{\today}
\maketitle

\section{Introduction}\label{sec:intro}

Flipping of spins due to the Barnett effect  \cite{Barnett} induced in a paramagnetic substrate by circularly polarized phonons has been recently demonstrated by Davies et al. \cite{Davies-Nature2024}. In essence, it corresponds to the generation of the effective magnetic field, ${\boldsymbol H}$,  due to a local twist via the Larmor theorem \cite{Larmor}. The latter states that in the coordinate frame rigidly coupled to the crystallographic axes of a solid, in which the localized spin state is formed, rotation at an angular velocity $\om$ is equivalent to the action of the magnetic field ${\boldsymbol H} = {\om}/\gamma$, with $\gamma$ being the gyromagnetic ratio associated with the spin. 

The Barnett effect and the reciprocal Einstein - de Haas effect \cite{Einstein1,Einstein2} were explored in the past as the mechanism of spin relaxation and decoherence due to interaction of spins with chiral phonons \cite{EC-PRB2002,EC-PRL2004,EC-DG-PRL2004,EC-DG-RS-PRB2005,Dornes-Nature2019,Tauchet-Nature2022}. These effects are also important for understanding of the conservation of angular momentum in experiments with nanocantilevers  \cite{Wallis-APL2006,JCG-PRB2009,Mori-PRB2020} and individual magnetic molecules that are free to rotate \cite{EC-PRL1994,EC-DG-PRB2010,DG-EC-PRB2021,Wernsdorfer-2015}. 

Numerous experiments reported flipping of spins in solids by applying short electric pulses \cite{Yang-Science2017}, laser  \cite{Xu-JMMM2022}, microwave  pulses \cite{Cai-PRB2013,Miyashita-PRL2023}, and surface acoustic waves \cite{Tejada-EPL2017,Camara-PRA2019}. From the quantum mechanical perspective, it can be understood by noticing that quantized phonons can carry angular momentum \cite{Zhang-PRL2014,DG-EC-PRB2015,Nakane-PRB2018,DG-EC-PRB2021}, which they can transfer to the spins. The Barnett and Enstein - de Haas effects provide a different angle from which this phenomenon can be viewed. 

Chiral phonons  generate an elastic twist with  angular velocity \cite{landautheory1986} ${\om} \sim \boldsymbol\nabla \times \dot{\boldsymbol u}$, where ${\boldsymbol u}$ is the phonon displacement field. For a phonon frequency $f=c/\lambda$, one can estimate $\dot{u}$ as $2\pi f u $, and its curl as $4\pi^2 f (u/\lambda)$,  with $\lambda$ being the wavelength and $c$ being the velocity of transverse sound waves. This gives $H = \Omega/\gamma\sim 4\pi^2(f/\gamma)(u/\lambda)$ for the effective magnetic field generated by such local elastic twists. For a powerful hypersound of frequency $f \sim\qty{e12}{\per\second}$, $c \sim \qty{3e3}{m/s}$, and $u \sim 0.1\lambda$, the field acting on the spin of the electron with  $\gamma =\qty{1.6e11}{\per\second\per\tesla}$ could be as high $4\pi^2(f/\gamma)(u/\lambda) \sim\qty{25}{T}$, which may explain large effective magnetic fields from high-frequency chiral phonons reported in $4f$ paramagnets \cite{Juraschek-PRR2022} and rare-earth halides \cite{Luo-Science2023}. 

On the contrary, the Barnett effect induced by a macroscopic mechanical rotation is always weak. Theoretical study on thin magnetic films and nanostructures performed by Bretzel et al. \cite{Bretzel-APL2009} has found that the rotational frequencies required to switch the magnetization in conventional materials would be beyond present experimental possibilities. Indeed, with  $\gamma \sim \qty{e11}{\per\second\per\tesla}$ for an electron spin, even a rotation as rapid as  $f = 10^3$ revolutions per second provides the effective field $2\pi f/\gamma$ less than $\qty{e-7}{T}$. A better candidate of magnetization by rotation would be fast-spinning microscopic grains in the interstellar medium, as has been proposed by the studies of meteorite paleomagnetism \cite{Weiss-SciAdv2021}.

This changes when one considers rotations induced by the defects of the crystal structure instead of global rotations. Recently, we have demonstrated \cite{EC-JS-PRB2025} that moving edge dislocations in a solid generate superfast local rotations. Since the spin quantum states are formed by the crystal field defined in the coordinate frame rigidly coupled with the axes of the crystal lattice, the spins feel local rotations as the effect of the magnetic field \cite{hehlinertial1990,EC-DG-RS-PRB2005}. We have shown that such effective magnetic fields due to fast-moving dislocations can easily flip individual spins embedded in the crystal lattice. 

In this paper, we study the effect of moving dislocations on the dynamics of the magnetization in a ferromagnetic film. This problem is significantly more involved than the problem solved in Ref.\ \cite{EC-JS-PRB2025} because it requires consideration of a system of many interacting spins coupled by the ferromagnetic exchange.  By solving numerically the Landau-Lifshitz equation in the lattice and laboratory coordinate frames in the presence of a moving edge dislocation, we demonstrate that such a dislocation, when assisted by an external magnetic field, can trigger the reversal of the magnetization of the film. 

The paper is organized as follows. The dynamics of the crystal lattice and interacting spins in the presence of a moving edge dislocation are studied in Section \ref{sec:dyn} in the lattice and laboratory coordinate frames for a spin Hamiltonian containing exchange interaction, magnetic anisotropy, and Zeeman interaction with the external magnetic field. Transformations from one coordinate frame to the other are discussed, and the corresponding equations of motion are derived. Numerical solution of the equations of motion is presented in Section \ref{mag-reversal}. Time evolution of the three-component magnetization of the film, generated by a moving dislocation, is obtained. Section \ref{conclusions} contains the discussion of our results and suggestions for experiments.

\section{Lattice dynamics}\label{sec:dyn}

\subsection{Effective magnetic field produced by a moving dislocation}
\begin{figure}\centering\includegraphics[width=0.8\linewidth]{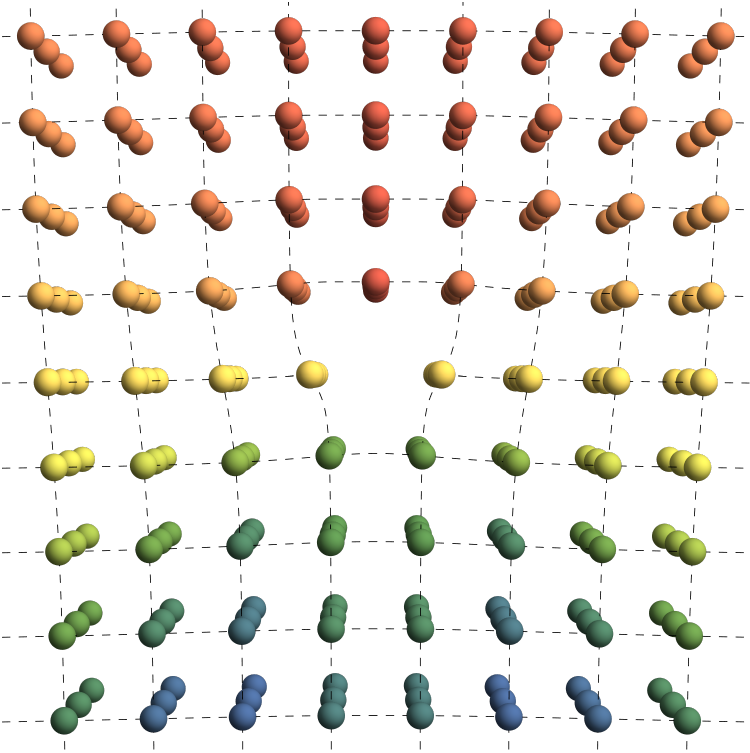}\caption{Deformed crystal lattice in the presence of an edge dislocation. The rotation of crystallographic axes away from the dislocation core is apparent.}\label{fig:lattice}\end{figure}

The deforming effect of edge dislocations on periodic lattices under the continuous approximation is well understood. The extra half-plane of atoms creates a displacement field around the dislocation core~\cite{burgerssome1939,landautheory1986}, resulting in a deformed lattice shown in Fig.~\ref{fig:lattice}. Dislocations are known to propagate, producing time-dependent deformations~\cite{eshelbyuniformly1949,nabarrotheory1967}. These deformations have a rotational component that gives rise to non-uniform angular velocities throughout the lattice~\cite{EC-JS-PRB2025}.

In the deformed lattice, atoms are at locations $\tilde{\boldsymbol x}_i=\boldsymbol x_i+\boldsymbol u(\boldsymbol x_i)$, where $\boldsymbol x_i$ are the original locations and $\boldsymbol u(\boldsymbol x)$ is the displacement field. From it, one can obtain the local rotation, $\boldsymbol\phi=\frac12\boldsymbol\nabla\times\boldsymbol u$, and the angular velocity, $\om=\frac12\boldsymbol\nabla\times\dot{\boldsymbol u}$, fields within the continuous elastic theory \cite{landautheory1986} as~\cite{EC-JS-PRB2025},
\begin{subequations}
\begin{equation}
	\boldsymbol\phi=-\frac{a}{2\pi}\frac{c_t^2}{c^2}\frac{\tilde\gamma_t^2(1-\gamma_t^2)}{\gamma_t}
	\frac{x-ct}{(x-ct)^2+y^2\gamma_t^2}\,\hat{\mathbf z}\label{eq:phi}
\end{equation}
and
\begin{equation}
	\om=-\frac{a}{2\pi}\frac{c_t^2}{c}\frac{\tilde\gamma_t^2(1-\gamma_t^2)}{\gamma_t}
	\frac{(x-c t)^2-\gamma_t ^2 y^2}{\left((x-c t)^2+\gamma_t ^2 y^2\right)^2}\,\hat{\mathbf z},\label{eq:omega}
\end{equation}
\end{subequations}
where
\begin{equation}
	\gamma_t=\sqrt{1-\beta^2},\quad\tilde\gamma_t=\sqrt{1-\beta^2/2},\quad\beta=c/c_t,
\end{equation}
$c$ is the dislocation propagation speed,
\begin{equation}
c_t=\sqrt{\frac{\mu}{\rho}}
\end{equation} is the speed of transversal waves in the material, $\rho$ its mass density, and $\mu$ its second Lam\'e coefficient (shear modulus). Notice that $\om=\dot{\boldsymbol\phi}$. Although these formulas are exact for an isotropic medium, they provide a good approximation for any symmetry of the crystal lattice away from the dislocation core.

The time-dependent deformations make individual atom rest frames not inertial, modifying their dynamics~\cite{hehlinertial1990,EC-DG-RS-PRB2005}. Effectively, they give rise to an additional coupling term in the Hamiltonian $-\hbar\boldsymbol S\cdot\om$. By comparison with the Zeeman coupling $-\gamma\hbar\boldsymbol S\cdot\boldsymbol H$, the local rotation has the same effect on the spin as the magnetic field ${\boldsymbol H} = \om/\gamma$. 

\subsection{Spin dynamics}
\subsubsection{Lattice frame Landau-Lifshitz equation}
We aim to describe the evolution of the individual atomic spins $\boldsymbol S_i$ as the dislocation traverses the lattice. We describe their evolution via the Landau-Lifshitz equation~\cite{chudnovskylectures2006}. The nature of this problem requires us to consider two types of coordinate frame: the lab frame and the local lattice frame at the location of an individual spin. Untilded quantities are in the lab frame. Tilded quantities with a $(k)$ superscript are in the local frame of the $k$-th spin.

The Landau-Lifshitz equation for each spin is naturally expressed in its own local frame as
\begin{equation}
	\frac{\mathrm d\st{i}{i}}{\mathrm d t}=\gamma\st{i}{i}\times\hefft{i}{i}-\alpha\gamma\st{i}{i}\times(\st{i}{i}\times\hefft{i}{i}),\label{eq:ll:latt}
\end{equation}
where
\begin{equation}
	\hefft{i}{i}=-\frac{1}{\gamma\hbar}\frac{\partial\tilde{\mathcal H}^{(i)}}{\partial\st{i}{i}}
\end{equation}
is the effective magnetic field, and $\tilde{\mathcal H}^{(i)}$ is the system Hamiltonian in the lattice frame of the $i$-th particle.

\subsubsection{Lattice frame hamiltonian}
We consider a ferromagnetic lattice with uniaxial anisotropy and exchange interactions subject to an external magnetic field and the propagating dislocation. Hence, the Hamiltonian in the coordinate frame of the $k$-th spin is
\begin{equation}
    \tilde{\mathcal H}^{(k)}=\tilde{\mathcal H}_\mathrm{ex}^{(k)}+\tilde{\mathcal H}_\mathrm{a}^{(k)}+\tilde{\mathcal H}_\mathrm{r}^{(k)}+\tilde{\mathcal H}_\mathrm{Z}^{(k)},
\end{equation}
where $\tilde{\mathcal H}_\mathrm{ex}^{(k)}$ is the exchange term, $\tilde{\mathcal H}_\mathrm{a}^{(k)}$ is the anisotropy term, $\tilde{\mathcal H}_\mathrm{r}^{(k)}$ is the dislocation-induced rotational term, and $\tilde{\mathcal H}_\mathrm{Z}^{(k)}$ is the Zeeman term, to be described below.

\paragraph{Exchange interactions}
The exchange Hamiltonian is
\begin{equation}
    \tilde{\mathcal H}_\mathrm{ex}^{(k)}=-\frac12\sum_{i,j}J_{ij}\st{i}{k}\cdot\st{j}{k}
\end{equation}
We take $J_{ij}=0$ except for nearest neighbors. Moreover, we take $J_{ij}=J$ to be a constant. This is justified because the main component of the deformation is rotational. Away from the dislocation core, it preserves the distances between the spins, upon which the exchange interaction depends.

\paragraph{Uniaxial anisotropy}
The magnetic anisotropy term couples to the spins via the Hamiltonian
\begin{equation}
    \tilde{\mathcal H}_\mathrm{a}^{(k)}=-\frac D2\sum_i\left(\nt{i}{k}\cdot \st{i}{k}\right)^2,\label{eq:h:a}
\end{equation}
where $D$ is the anisotropy constant and $\nt{i}{k}$ is the anisotropy unit vector at the position of the $i$-th spin. Since the anisotropy is a local property that follows the deformations in the lattice, the vector $\nt{i}{i}$ is a constant.

\paragraph{Zeeman term}
While the external field is most naturally expressed in the lab frame as a constant vector, we can formally write 
\begin{equation}
    \tilde{\mathcal H}_\mathrm{Z}^{(k)}=-\gamma\hbar\sum_i\st{i}{k}\cdot\tilde{\boldsymbol H}^{(k)}
\end{equation}
in the rotating frame.

\paragraph{Rotation term}
The deformations introduced by the dislocation produce an effective magnetic field $\om_i/\gamma$ at the location of the $i$-th spin, with $\om_i=\om(\boldsymbol x_i)$. Thus, we write
\begin{equation}
    \tilde{\mathcal H}_\mathrm{r}^{(k)}=-\hbar\sum_i\st{i}{k}\cdot\omt{i}{k}.
\end{equation}

\subsubsection{Lattice frame effective field}
We use the above Hamiltonian to obtain the effective magnetic field for the $i$-th spin:
\begin{multline}
	\hefft{i}{i}=\frac{1}{\gamma\hbar}\left(J\sum_j \st{j}{i}+D\left(\st{i}{i}\cdot \nt{i}{i}\right)\nt{i}{i}\right.\\\left.\vphantom{\sum_j}+\hbar\omt{i}{i}+\gamma\hbar\tilde{\mathbf H}^{(i)}\right),\label{eq:heff:latt}
\end{multline}
where the index $j$ sum spans over the nearest neighbors of the $i$-th spin.

\subsubsection{Lab frame Landau-Lifshitz equation}
Local rotation in the $xy$ plane by an angle $\phi=\boldsymbol\phi\cdot\hat{\mathbf z}$ establishes a relation between the basis vectors in the two frames as
\begin{equation}
	\begin{aligned}
		\bvt{x}{i} & =  \cos\phi_i^{(i)}\,\bv{x}+\sin\phi_i^{(i)}\,\bv{y},\\
		\bvt{y}{i} & = -\sin\phi_i^{(i)}\,\bv{x}+\cos\phi_i^{(i)}\,\bv{y},\\
		\bvt{z}{i} & =  \bv{z}.
	\end{aligned}\label{eq:basis}
\end{equation}
Since this rotation does not affect the $z$ components of vectors, we drop any frame-related labels in $\boldsymbol\phi$ and $\om$.

The transformation \eqref{eq:basis} induces a rotation $\tilde{\boldsymbol v}^{(i)}=R_i\boldsymbol v$ of the components of all vectors involved, with the rotation matrix
\begin{equation}
	R_i=\begin{pmatrix}
		\cos\phi_i & \sin\phi_i & 0\\
		-\sin\phi_i & \cos\phi_i & 0\\
		0        & 0        & 1
	\end{pmatrix}.
\end{equation}

With this in mind, the dynamical equation for the $i$-th spin \eqref{eq:ll:latt} may be rewritten in the lab frame as 
\begin{equation}
\frac{\mathrm d\s{i}}{\mathrm d t}=	\gamma\s{i}\times\heff{i}-\alpha\gamma\s{i}\times(\s{i}\times\heff{i})
-R^{-1}_i\dot R_i\s{i}.
\end{equation}
Since
\begin{equation}
	R^{-1}_i\dot R_i\boldsymbol S=\dot\phi_i\begin{pmatrix}0&1&0\\-1&0&0\\0&0&0\end{pmatrix}\s{i}=\dot\phi_i\s{i}\times\bv{z}=\s{i}\times\om_i,
\end{equation}
we may rewrite the lattice frame Landau-Lifshitz equation as 
\begin{equation}
	\frac{\mathrm d{\s{i}}}{\mathrm d t}=\gamma{\s{i}}\times(\heff{i}-\om_i/\gamma)-\alpha\gamma\s{i}\times(\s{i}\times\heff{i}),\label{eq:ll:lab}
\end{equation}
effectively canceling the rotation-induced term in the gyroscopic term but not in the dissipation term. 

This is the essence of the Barnett effect: spins relax toward the direction of the effective field created by the exchange interaction, magnetic anisotropy, the external field, and the effective field due to rotation. 

\subsubsection{Dimensionless equation}
The complexity of the system requires careful choice of numerical techniques. In order to remove the dependence on physical units, we now turn to the dimensionless form of the problem. To do so, we introduce a time constant $t_0$, and a dimensionless time $\tau=t/t_0$.  This allows us to define dimensionless exchange and anisotropy constants as $\theta=t_0 J/\hbar$ and $\nu={t_0 D}/{\hbar}$, the dimensionless external field as $\boldsymbol h=\gamma t_0\boldsymbol H$, and the dimensionless angular velocity as $\boldsymbol\omega_i=t_0\om_i$.
The final form of the equation is
\begin{multline}
	\frac{\mathrm d\s{i}}{\mathrm d\tau}=
	\s{i}\times\left(\theta\sum_j\boldsymbol S_j+\nu(\s{i}\cdot\mathbf n_i)\mathbf n_i+\boldsymbol h\right)\\
	-\alpha \s{i}\times\left[\s{i}\times\left(\theta\sum_j\boldsymbol S_j+\nu(\s{i}\cdot\mathbf n_i)\mathbf n_i+\boldsymbol h+\boldsymbol\omega_i\right)\right].\label{eq:ll:dimless}
\end{multline}

\section{Magnetization reversal by a moving dislocation}\label{mag-reversal}
\subsection{Numerical analysis}
In this work, we illustrate the feasibility of reversing a film's magnetization by the passage of a moving dislocation by choosing a hexagonal lattice with 10191 sites, and a nearly-square shape, using the fact that away from the dislocation core, our model of the dislocation developed within isotropic elastic theory provides a reasonably good approximation for the lattice of any symmetry. We further constrain the physical parameters of the system (exchange and anisotropy constants  and speed of sound) by focusing on cobalt as a magnetic material. 

In order to choose a time constant relevant for the problem, we consider the Debye frequency for cobalt, $f_D=\qty{9.3e12}{Hz}$, which yields $t_0=f_D^{-1}=\qty{1.07e-4}{ns}$. The speed of transversal waves may be obtained for cobalt from $\rho=\qty{8.9e3}{kg/m^3}$ and $\mu=\qty{75}{GPa}$ as $c_t\approx\qty{2.9e3}{m/s}$. We will consider values of $\beta=c/c_t\approx1$.

The anisotropy constant $D$ from \eqref{eq:h:a} is related to the anisotropy energy density as $D=2K_{u1} V$, where the volume per atom is $V=m/\rho\approx\qty{11}{\angstrom^3}$. Taking $K_{u1}=\qty{4.1e5}{J\per\meter\cubed}$~\cite{krishnanfundamentals2020}, we get a dimensionless anisotropy constant $\nu\approx\num{9.19e-3}$. The exchange energy between the nearest neighbors in a cobalt hexagonal lattice is of the order of $J\approx\qty{3.21e-2}{eV}$~\cite{pajdaabinitio2001}, which yields a dimensionless exchange constant of $\theta\approx\num{5.25}$. We consider a damping parameter $\alpha\approx\numrange{e-3}{e-2}$~\cite{ooganemagnetic2006,ebertabinitio2011}. We chose an anisotropy easy axis along the $z$ axis, such that $\nt{i}{i}=\left<0,0,1\right>$.

With these choices, we evolve the Landau-Lifshitz equation \eqref{eq:ll:dimless} for the three components of the spin of each atom, using a fourth-order Runge-Kutta parallel evolver that uses pre-computed values of the rotation fields $\boldsymbol\phi$ and $\boldsymbol\omega$ at the locations of the particles at the desired time instants. Below, we characterize the meaning of the magnetization switch and present numerical results to illustrate this phenomenon.

\subsection{Equilibria}
A successful switch in the magnetization shall bring the individual spins from one stable equilibrium ($\dot{\boldsymbol S_i}=\boldsymbol0$ and negative eigenvalues of the Jacobian) of the system to another. Without an external magnetic field, two such equilibria exist: $\s{i}=\left<0,0,\pm1\right>$, for all $i$. The application of an external field breaks the symmetry between them: for an upward-pointing field ($h_z>0$), the $s_{i,z}<0$ equilibrium becomes metastable for small fields, and unstable for large fields. We consider upward-pointing fields that preserve its metastability, and take that configuration as the initial state of the system, with the expectation that the dislocation will switch the system towards the $s_{i,z}>0$ stable equilibrium.

\begin{figure}\centering\vspace{1em}\includegraphics[width=0.8\linewidth]{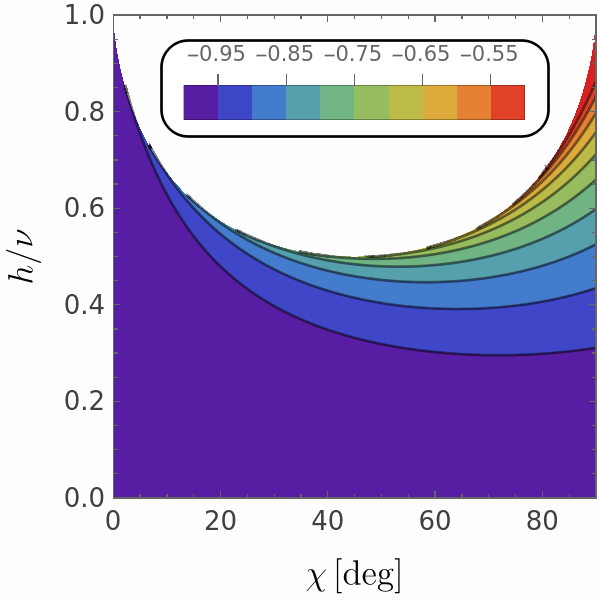}\caption{Existence of meta-stable equilibria opposing the external field, for $h_z>0$. The equilibrium value of $s_{i,z}$ is shown in color.}\label{fig:equilibria}\end{figure}

Describing the external field in spherical coordinates as $\boldsymbol h=\left<h\sin\chi\cos\varphi,h\sin\chi\sin\varphi,h\cos\chi\right>$, the existence of equilibria is independent of $\varphi$. In Fig.~\ref{fig:equilibria} we show the values of $h/\nu$ and $\chi$ for which meta-stable equilibria (opposing the field) exist. It is worth noting that a purely vertical field ($\chi=\pi$) would result in a static situation due to the symmetry of the system.

Below, we consider a magnetization reversal from a large $-s_{i,z}$ to a large $s_{i,z}$ with a small field angle $\chi$.

\subsection{Results}
\begin{figure}\includegraphics[width=\linewidth]{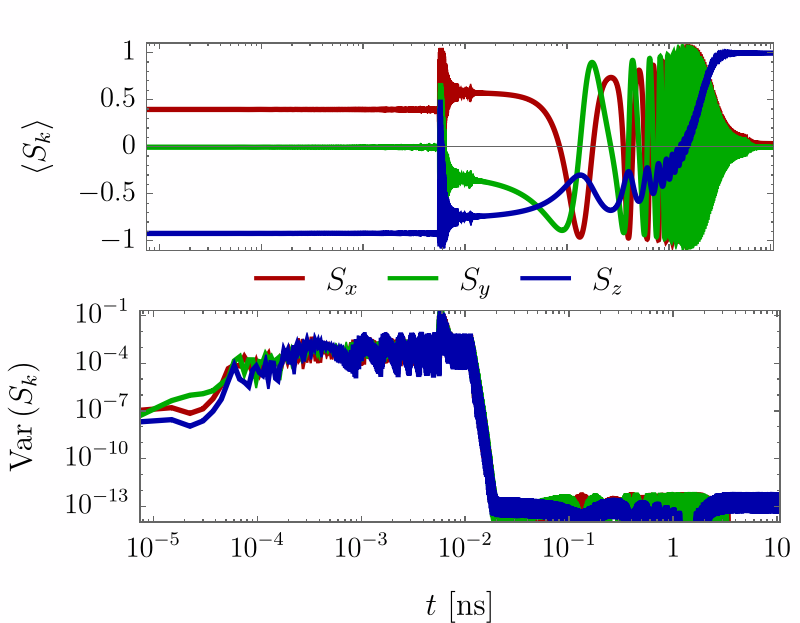}\caption{Full evolution of the lattice.}\label{fig:avg:full}\end{figure}

For illustration, we choose the magnetic parameter values for cobalt mentioned above ($\nu=\num{9.19e-3}$, $\theta=\num{5.25}$ and $\alpha=0.01$), together with a dislocation speed $\beta=0.99995$ and a magnetic field with $h=\num{7e-3}$, $\chi=\qty{5}{\degree}$ and $\varphi=\qty{0}{\degree}$, such that $H\approx\qty{0.69}{T}$. The uniformly magnetized lattice starts at a meta-stable equilibrium with spins $\boldsymbol S_i(t=0)\approx\left<0.4011,0,-0.9160\right>$, about $\qty{23}{\degree}$ away from the negative $z$ axis. In Fig.~\ref{fig:avg:full} we show the full evolution of the average magnetization and its variance throughout the lattice, which ends in a reversed magnetization.

\begin{figure}\includegraphics[width=\linewidth]{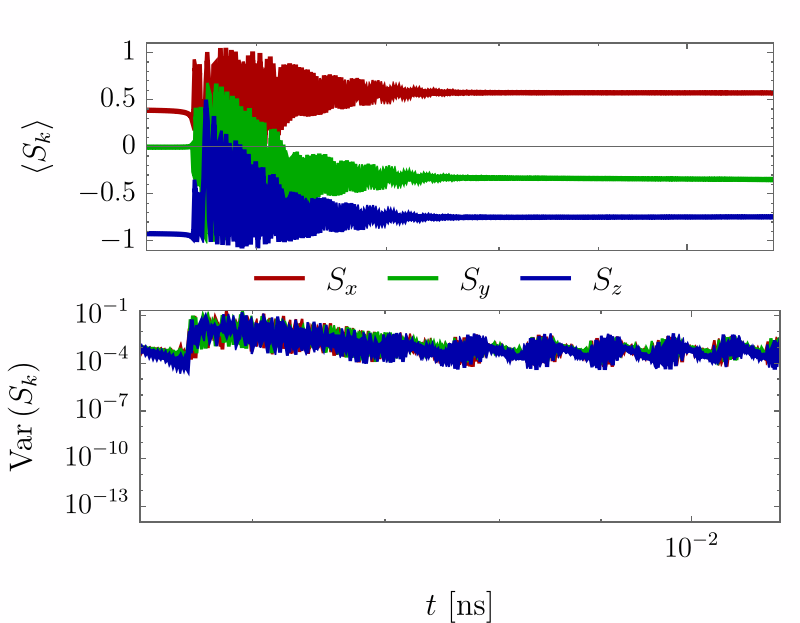}\caption{Evolution of the lattice as the dislocation passes.}\label{fig:avg:passagedetail}\end{figure}

The evolution is first characterized by a slow movement away from the meta-stable equilibrium, as the dislocation enters the lattice, during the first $\sim\qty{5}{ps}$.
Then, the dislocation passes through the lattice for about \qty{10}{ps}, producing rapid and disordered evolution (Fig.~\ref{fig:avg:passagedetail}) of the spins near the dislocation core, which quickly propagates away from it. The dislocation brings the spins away from the meta-stable equilibrium, as it is clearly visible in the shift of $\left<S_z\right>$ before and after its passage. 

\begin{figure}\includegraphics[width=\linewidth]{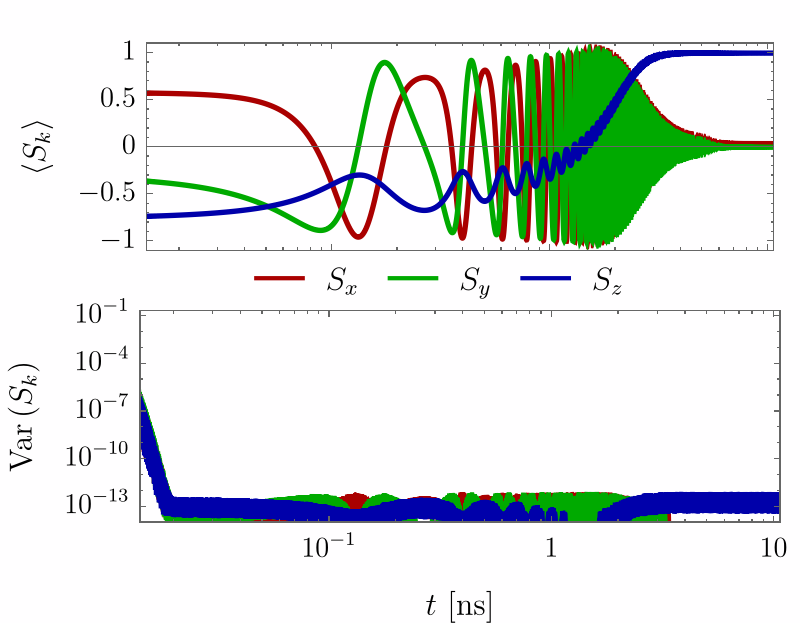}\caption{Evolution of the lattice after the dislocation has passed, as it relaxes towards the stable equilibrium.}\label{fig:avg:relaxation}\end{figure}

After the dislocation exits the lattice, the spins align again, producing an ordered state shortly after (about another \qty{10}{ps} after the dislocation has passed, at a total time of $t\approx\qty{2e-2}{ns}$). At this point, the rotation-free gyroscopic term and the damping term start driving oscillations of the spins with the time scale of about  $\qty{0.04}{ns}$ ($\qty{25}{GHz}$). Eventually, the spins fully relax on a time scale between $1$ns and $10$ns (see Fig.~\ref{fig:avg:relaxation}). It is interesting to notice that in this complex process, the well-defined and well-seen reversal of the magnetization occurs in a very short time of just several picoseconds.

\section{Conclusions}\label{conclusions}
The magnetization curve of most magnets exhibits magnetic hysteresis due to long-living metastable states created by the magnetic anisotropy.  In a stressed material, short displacements of dislocations due to thermal fluctuations may add to the slow magnetic relaxation due to thermal motion of domain walls in a random potential landscape formed by crystal defects. Here, we have studied a different situation that occurs when a dislocation traverses a noticeable distance inside a ferromagnet at a high speed. 

It has been well established that under a large elastic stress, dislocations can move at a speed comparable to the speed of sound \cite{Benat2021}. A shock stress that produces a highly dense, rapidly-developing dislocation structure can be achieved by mechanical means, or by a short voltage pulse in a ferroelectric material, or by an ultrashort laser pulse \cite{Matsuda-JAP2014}. The influence of the mechanical stress on the magnetization has been commonly attributed to the magneto-elastic coupling; the effect of propagating dislocations has been largely overlooked.

Here, we have demonstrated that due to the Barnett effect, an individual dislocation moving at a high speed can facilitate the decay of a metastable magnetic state created by the external magnetic field. The switching of the magnetization occurs due to the large effective magnetic fields generated by the elastic twists around the dislocation core. In a ferromagnetic film of a small area, the reversal of the magnetization can be very fast; it took several picoseconds in our numerical experiment on a lattice containing $10^5$ spins. 

Manipulating magnetization by voltage has been a paradigm of contemporary applied magnetism \cite{Fert-RMP2024}. It would be interesting to see if the Barnett effect could be used for that purpose in, e.g., nanoscale multiferroics where a very rapid shear deformation can be induced by the electric field \cite{EC-RJ-JAP2015}. This could open a path to the utilization of the Barnett effect at the nanoscale for electric  manipulation of magnetic memory units. 

In larger systems, the effect of dislocations on the magnetization reversal studied in this paper must be greatly enhanced by an avalanche of rapidly propagating dislocations formed spontaneously on the approach to plastic deformation by crossing the ultimate tensile strength of the material. Observation of this effect requires rather simple experiments that we hope will be performed in the future. 

\section{Acknowledgements}
This work has been funded by the U.S. Air Force Office of Scientific Research (AFOSR) through Grant No. FA9550-24-1-0290.

\end{document}